\documentclass[letterpaper, 10 pt, conference]{ieeeconf}  % Comment this line out if you need a4paper

\usepackage{graphicx,keyval,trig,url} %,biblatex}
\usepackage[T1]{fontenc}     %LaTeX Font Warning: Font shape `OMS/cmtt/m/n' undefined 
\usepackage[utf8]{inputenc}	%apostrophes appear normally
\usepackage{graphicx}

\usepackage{amsthm}
\usepackage{amssymb}
\usepackage{amsmath}
\usepackage{mathtools}
\usepackage[mathscr]{eucal}
\usepackage{lipsum}
\usepackage{cuted} % environment changes the formatting from two-column to one-column to better accommodate very long equations
\usepackage{blindtext}
\usepackage{color}
\usepackage{dsfont}

\usepackage{epstopdf}
\usepackage{bm}   %turn the letter into bold in the math mode

\usepackage{color}
\usepackage{xcolor}

\usepackage{multirow}
\usepackage{longtable}

\usepackage{mathtools}
\usepackage[colorinlistoftodos]{todonotes}

\usepackage{algorithm}
\usepackage{algpseudocode}

\usepackage{xcolor}
\usepackage{import}

\usepackage[normalem]{ulem} % strikeout commands

\usepackage{subfigure}
\usepackage{blindtext}
\usepackage{tabularx}

\usepackage{color}
\definecolor{ajg}{RGB}{0,0,0}
\definecolor{zx}{RGB}{0,0,0}
\definecolor{akp}{RGB}{0,0,0}

\IEEEoverridecommandlockouts                              % This command is only needed if 
\title{\LARGE \bf An Economic Model Predictive Control Approach for Load Mitigation on Multiple Tower Locations of Wind Turbines}

\author{Zhixin Feng, Alexander J. Gallo, Yichao Liu, Atindriyo K. Pamososuryo,\\ Riccardo M.G. Ferrari and Jan-Willem van Wingerden  % <-this % stops a space
\thanks{
    \textcolor{zx}{This work is partially supported by the WATEREYE project, which
    is funded by European Innovation and Networks Executive
    Agency under the European Union’s Horizon 2020 research and innovation program under grant agreement no. 851207, and was also supported by the AIMWIND project, funded by the Research Council of Norway under grant no. 312486.}} 
\thanks{
    Authors are with the Delft Center for Systems and Control, Delft University of Technology, Mekelweg 2, 2628 CD Delft, The Netherlands.
    {\tt\small \{Z.Feng-2, A.J.Gallo, Y.Liu-17, A.K.Pamososuryo, R.Ferrari, J.W.vanWingerden\}@tudelft.nl}.}}

\begin{document}

\maketitle
\thispagestyle{empty}
\pagestyle{empty}

%%%%%%%%%%%%%%%%%%%%%%%%%%%%%%%%%%%%%%%%%%%%%%%%%%%%%%%%%%%%%%%%%%%%%%%%%%%%%%%%
\begin{abstract}
The current trend in the evolution of wind turbines is to increase their rotor size in order to capture more power. This leads to taller, slender and more flexible towers, which thus experience higher dynamical loads due to the turbine rotation and environmental factors. It is hence compelling to deploy advanced control methods that can dynamically counteract such loads, especially at tower positions that are more prone to develop cracks or corrosion damages. Still, to the best of the authors' knowledge, little to no attention has been paid in the literature to load mitigation at multiple tower locations. Furthermore, there is a need for control schemes that can balance load reduction with optimization of power production.
In this paper, we develop an Economic Model Predictive Control (eMPC) framework to address such needs. 
First, we develop a linear modal model to account for the tower flexural dynamics. 
Then we incorporate it into an eMPC framework, where the dynamics of the turbine rotation are expressed in energy terms. This allows us to obtain a convex formulation, that is computationally attractive.
\textcolor{zx}{Our control law is designed to avoid the ``turn-pike'' behavior and guarantee recursive feasibility.}
%\textcolor{ajg}{Our control law is designed such that the so-called ``turn-pike'' behavior %typical of solving economic objectives is avoided, and guaranteeing recursive feasibility.}
\textcolor{ajg}{We demonstrate the performance of the proposed controller on a 5MW reference WT model:}
% The performance of the proposed solution is demonstrated on a 5MW reference wind turbine model.
the results illustrate that the proposed controller is able to reduce the tower loads at multiple locations, without significant effects to the \textcolor{ajg}{generated power.}
% power generation.

\end{abstract}

%%%%%%%%%%%%%%%%%%%%%%%%%%%%%%%%%%%%%%%%%%%%%%%%%%%%%%%%%%%%%%%%%%%%%%%%%%%%%%%%

\section{Introduction}\label{sec:intro}
\noindent 
% In past decades, 
Wind energy has \textcolor{ajg}{recently} received increasing attention in the international energy market.
In 2020, 90 GW of new wind power capacity was deployed, contributing to a global growth of 53\% compared to 2019~\cite{council2021gwec}.
Such growth is partially driven by the increasing physical dimension of wind turbines (WTs), which allows for more wind power to be captured.
However, higher fatigue loads on the increasingly flexible WT towers are also experienced as a downside of this trend.
Extra attention, therefore, needs to be paid to  mitigate the structural loads while keeping power production minimally affected, as these objectives are often competitive.
From the control engineering standpoint, this urges for the employment of advanced controllers, capable of addressing the power regulation and load mitigation trade-off.
In the literature, a number of control algorithms that are able to cater for the aforementioned trade-off have been proposed. 
For instance, a quasi-linear parameter varying model predictive control scheme was used in~\cite{Mulders2020} and an adaptive gain scheduling proportional–integral (PI) control was proposed in~\cite{Lara2021}. 
In addition, the economic model predictive control (eMPC) framework was proposed to optimize the aforementioned trade-off~\cite{Shaltout2017},~\cite{AtinJoP2022}.
eMPC is a control paradigm that has been introduced in the past decade to include economic considerations in the objective function of predictive controllers, rather than tracking reference points~\cite{rawlings2012fundamentals}.
% ; recent reviews on nonlinear economic MPC can be found in \cite{faulwasser2018economic}.
The control methods mentioned above mainly account for fatigue loads at the tower bottom location. 
However, critical damage can be caused by fatigue loads on other tower locations as well, where cracks or serious corrosion can occur.
Thus, the reduction of fatigue loads at more than one location along the WT's tower is of importance.
Still, to the best of the authors' knowledge, there are no contributions in literature addressing the design of a controller capable of optimizing the trade-off of power generation and the reduction of loads at multiple tower locations.
\textcolor{zx}{Additionally, \textit{turnpike} behavior is a common feature of finite-horizon eMPC~\cite{rawlings2012fundamentals,faulwasser2018economic}.} \textcolor{zx}{This undesired behaviour refers to the fact that the optimizer may drive the nominal system away from the optimal steady state at the end of the prediction horizon.}
%\textcolor{ajg}{The reason is that, since the horizon is finite, the optimization problem does not %consider anything that occurs after it: as such, all possible economic value is extracted from the system %before the final time.
%\textcolor{ajg}{Specifically, for WTs, as pointed out %in~\cite{gros2013economic}, this behavior is equivalent to the controller %converting as much kinetic energy as possible into generated power towards %the end of the horizon, without including any information as to what occurs %past the prediction horizon.} 
\textcolor{zx}{Although addressed in \cite{gros2013economic} by including a suitably defined term in the objective function of the eMPC-based controller, analysis of appropriate solutions to avoid the turnpike behavior is lacking from a large number of works in the literature addressing eMPC for WTs.}

% ALEX TRYING TO REWORD CONTRIBUTION SECTION
 In this paper, we develop an eMPC-based controller that simultaneously reduces loads at multiple WT tower locations and maximizes power generation.
%  Our main contribution is that we leverage modal analysis to include a higher order approximation of the tower fore-aft vibrations dynamics \cite{gawronskiAdvancedStructuralDynamics2004}.
%  This allows us to predict the state of the WT more accurately, thus achieving higher performance with respect to existing control strategies.
%  Additionally, and differently to a large number of works in the literature addressing eMPC for WTs, we include a terminal constraint in the eMPC law, based on the optimal steady state.
%  This not only ensures that optimal trajectories do not exhibit turnpike behavior, but also improves the control performance.
%  Extensive simulation results are presented, where our proposed controller is applied to the National  Renewable Energy Laboratory (NREL)'s 5MW reference wind turbine model~\cite{jonkmanDefinition5MWReference2009}.
Our main contributions are
\begin{itemize}
    \item we leverage modal analysis to include a higher order approximation of the tower fore-aft flexural dynamics~\cite{gawronskiAdvancedStructuralDynamics2004};
    \item we improve performance by introducing a terminal constraint in the eMPC law, based on the optimal steady state, which avoids the turn-pike behavior;
    \item we \textcolor{zx}{apply the proposed controller to the National  Renewable Energy Laboratory (NREL)'s 5MW reference WT model~\cite{jonkmanDefinition5MWReference2009} and present the extensive results.}
    %present extensive simulation results, where our proposed controller is %applied to the National  Renewable Energy Laboratory (NREL)'s 5MW %reference wind turbine model~\cite{jonkmanDefinition5MWReference2009}.
\end{itemize}

The higher order flexural model allows us to predict the state of the WT more accurately, thus achieving higher performance with respect to existing control strategies.
Additionally, the inclusion of a terminal constraint ensures that optimal trajectories do not exhibit \textit{turnpike} behavior, while also improving control performance. 
%\textcolor{ajg}{Turnpike behavior is a common feature of finite-horizon eMPC, as pointed out in %\cite{rawlings2012fundamentals,faulwasser2018economic}.
%This undesired behaviour is caused by the fact that, to optimize economic objectives, the optimizer of \eqref{eq:FHOCP} may drive the nominal system away from the optimal steady state at the end %of the prediction horizon.
%Including a terminal constraint guarantees that the system does not depart from the optimal steady state value at the end of the prediction horizon.}
%\textcolor{zx}{By including a terminal constraint, the proposed controller guarantees that the system does %not depart from the optimal steady state value at the end of the prediction horizon.}
The effectiveness of the proposed controller is compared to other eMPC-based solutions from the literature, specifically: one with single-location tower load reduction~\cite{Shaltout2017}, and one in which tower loads are not included in the controller objective function~\cite{Hovgaard2015}.
Analysis of the effects of the prediction horizon length is also given, highlighting the need to balance performance and computational complexity. In this respect, we note how including a terminal constraint further improves performances at a negligible computational cost.

The remainder of this paper is organized as follows: in Section~\ref{sec:model} the system dynamic model is defined. Section~\ref{sec:eMPC} formalizes the eMPC framework for loads reduction on multiple tower locations. In Section~\ref{sec:simulation}, case studies are carried out to numerically demonstrate the proposed controller. Finally, conclusions are drawn in Section~\ref{sec:conclusions}.

\section{Definition of the System Dynamic Model}\label{sec:model}
\noindent Let us start by introducing the dynamical model of the WT used in this paper; specifically, we utilize a single rotational model to describe the dynamics of the turbine drive train, and a multi-mode model to approximate the tower vibration. 
%For the former, we use the simplified single rotational mass model~\cite{Hovgaard2015}.
%Following this, we introduce a nonlinear transformation of coordinates that allows the definition of linear dynamics with convex constraints, to be used in the eMPC-based controller designed in Section~\ref{sec:eMPC}.

% We consider the dynamics of two subsystems of the turbine: those related to the drive train, composed of the rotor, gearbox and generator, and the tower vibration.
% For the former, we use the simplified single rotational mass model~\cite{Hovgaard2015}, while for the latter we define a multi-mode tower vibration model.

\subsection{Drive Train Dynamics}
\noindent Let $\omega_\mathrm{g}$ denote the generator angular speed. 
Then the single-order model of the drive train dynamics is formulated as:
\begin{equation}
     \label{DriveTrainDynamics}
    \dot{\omega}_{\mathrm{g}}=\frac{1}{J}\left[\frac{1}{G} T_{\mathrm{r}}-T_{\mathrm{g}}\right] 
    \, ,
\end{equation}
where $T_{\mathrm{r}}$, $T_{\mathrm{g}}$ and $G\geq1$ are the rotor torque, generator torque and gearbox ratio, respectively; $J = J_{\mathrm{g}} + J_{\mathrm{r}}/G^2$ represents the equivalent inertia at the generator shaft, where $J_{\mathrm{r}}$ and $J_{\mathrm{g}}$ are, respectively, the rotor and generator inertia; and, supposing a stiff rotor shaft, the rotor speed is
%We suppose the gearbox to be perfectly stiff, and thus there are no torsional %dynamics linking the rotor and generator speed, which are related by
$\omega_\mathrm{r} = \omega_\mathrm{g}/G$.
The rotor torque $T_\mathrm{r}$ is defined by:
\begin{equation}
    %\label{AerodynamicTorqueCalculation}
T_{\mathrm{r}}=\frac{1}{2 \omega_{\mathrm{r}}} \rho A C_{\mathrm{p}}(\lambda, \beta) v_{\mathrm{w}}^{3}
\, ,
\end{equation}
where $\rho$ is the air density; $A$ is the rotor swept area; $\beta$ is the blade pitch angle; $v_\mathrm{w}$ is the wind speed; and 
$\lambda$ represents the tip-speed ratio, defined as:
\begin{equation}
    %\label{tip-speedRatioCalculation}
    \lambda = \frac{\omega_{\mathrm{r}} D_{\mathrm{r}}}{2v_{\mathrm{w}}} = \frac{\omega_{\mathrm{g}} D_{\mathrm{r}}}{2Gv_{\mathrm{w}}} 
    \, ,
    \label{Eq:lambda}
\end{equation}
where $D_{\mathrm{r}}$ is the rotor diameter.
Finally, $C_\mathrm{p}$ is the nonlinear power coefficient, specific to each WT, which is derived via experiments or steady-state simulations.
For the NREL's 5MW WT \cite{jonkmanDefinition5MWReference2009} considered in this work, look-up tables for $C_\mathrm{p}$ have been derived for control design.

The aerodynamic power extracted from the wind, $P_\mathrm{r}$, and the generator power, $P_\mathrm{g}$, are defined as:
\begin{align}
    \label{Eq:Pr}
    P_{\mathrm{r}}&=T_{\mathrm{r}} \omega_{\mathrm{r}}=\frac{1}{2} \rho A C_\mathrm{p}(\lambda, \beta) v_{\mathrm{w}}^{3} \, ,\\
     P_{\mathrm{g}}&=\eta_{\mathrm{g}} T_{\mathrm{g}} \omega_{\mathrm{g}}\label{eq:Pg}
    \, ,
\end{align}
where $\eta_{\mathrm{g}}$ is the generator efficiency. 
As is standard in this modeling framework, we consider $\omega_\mathrm{g}$ as the state variable of the system, while $\beta$ and $T_\mathrm{g}$ are controllable inputs, and $v_\mathrm{w}$ is an uncontrolled input to the system.

%\begin{remark}
%    We note explicitly that, although not apparent from \eqref{DriveTrainDynamics}, the dynamics of the wind turbine are nonlinear.
%    Indeed, $\omega_\mathrm{g}$ appears in the definition of $T_\mathrm{r}$ through the nonlinear coefficient $C_\mathrm{p}$.
%    Another source of nonlinearity exists in the formulation of $P_{\mathrm{g}}$ in \eqref{eq:Pg} being a bilinear function of $T_{\mathrm{g}}$ and %$\omega_{\mathrm{g}}$.
%    $\hfill\triangleleft$
%\end{remark}

For proper operation of the WT, the following state, input, and output constraints must be guaranteed~\cite{Hovgaard2015}:
 %the state $\omega_\mathrm{g}$ must satisfy
% \begin{equation}\label{eq:cstr:state}
%     \omega_{\mathrm{g,min}}\leq \omega_\mathrm{g}\leq \omega_{\mathrm{g,max}}\,,
% \end{equation}
%similarly, the generator torque and pitch angle have upper and lower bounds:
\begin{align}%\label{eq:cstr:input}
        \omega_{\mathrm{g,min}} &\leq \omega_\mathrm{g}\leq \omega_{\mathrm{g,max}} \,, \label{eq:cstr:state}\\
        0 &\leq T_\mathrm{g} \leq T_{\mathrm{g,max}} \,, \label{eq:cstr:inputTg}\\
        \beta_{\mathrm{min}} &\leq \beta \leq \beta_{\mathrm{max}} \,, \label{eq:cstr:inputBeta}\\
        0 &\leq P_\mathrm{g} \leq P_\mathrm{g,rated} \,. \label{eq:cstr:output}
\end{align}
%Finally, appropriate definition of the control law must ensure that the generator power does not exceed %some rated values, \textit{i.e.}, that $P_\mathrm{g}$ satisfies:
% \begin{equation}
%     0\leq P_\mathrm{g}\leq P_\mathrm{g,rated}\,.
% \end{equation}

\subsection{Multi-mode Tower Vibration Model}

\begin{figure}
    \centering
    \begin{minipage}[c]{0.42\columnwidth}
       \centering
       \subfigure[Modal analysis model]{
    \centering
    \label{fig:TowerMAa}
    \includegraphics[width=\textwidth]{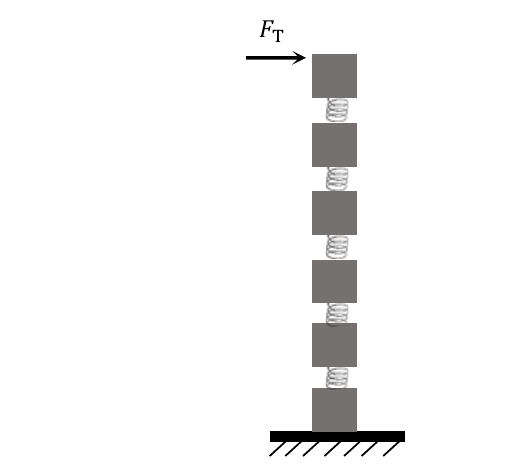}}
    \end{minipage}
    \hfill
    \begin{minipage}[c]{0.42\columnwidth}
       \centering
       \subfigure[Mode Shapes]{
    \centering
    \label{fig:TowerMAb}
    \includegraphics[width=\textwidth]{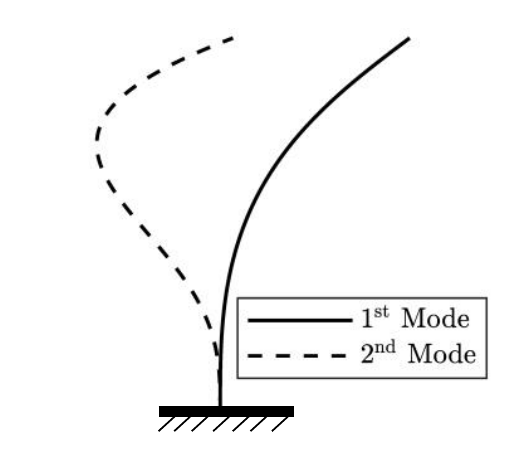}}
    \end{minipage}
    \caption{Tower vibration model and its mode shapes}
    \label{fig:TowerMA}
\end{figure}

\noindent
Having presented the dynamical model of the WT drive train, let us now define our modeling of the WT tower vibrations.
%\textit{i.e.}, the description of the dynamics of the vibration displacement of the tower at different heights, driven by the aerodynamic loads of the structure. 
Specifically, we are interested in the tower fore-aft vibrations.

Let $F_{\mathrm{T}}$ denote the thrust force. As shown in Fig.~\ref{fig:TowerMAa}, the tower structure can be modeled as a multiple mass-spring-damper system \textcolor{zx}{which is fixed at the bottom}. 
Here we approximate \textcolor{zx}{the tower vibration} dynamics via modal analysis~\cite{gawronskiAdvancedStructuralDynamics2004}, through which the vibration can be decomposed into different contributing modes. In Fig.~\ref{fig:TowerMAb} we show the mode shapes of the first two contributing modes. \textcolor{zx}{Let $\mathbf{\Phi}\in\mathbb{R}^{N_\mathrm{d} \times N_\mathrm{m}}$ represent the mode-shape matrix, with $\Phi_i = [\phi_{1,i},\dots, \phi_{N_d,i}]^\top$ being the $i^\text{th}$ column of the matrix and $N_\mathrm{d}$ the number of degrees of freedom.}
%
% \noindent Having presented the dynamical model of the wind turbine drive train, in the following we describe the multi-mode model of the tower vibration which is included in the control design, driven by the aerodynamic loading of the structure.
% As shown in Fig.\ref{fig:TowerMAa}-\ref{fig:TowerMAb}, the tower can be modeled as a multiple mass-spring system. 
% Based on the modal analysis~\cite{gawronskiAdvancedStructuralDynamics2004}, the vibration can be decomposed into  several contributing modes. 
% In Fig.\ref{fig:TowerMAc} we show the mode shapes of the first two contributing components. 
%
%
Let then $N_l$ denote the number of locations: then  $z_l \in [0,1], l \in \{1,\dots,N_l\}$ represent the \textit{normalized} heights, \textcolor{zx}{with $z_1 = 1$}.
To exploit the modeling modal framework, we must be able to relate $\mathbf{x}_{\mathrm{p}} = [x_{\mathrm{p},1},\dots,x_{\mathrm{p},N_l}]^\top$, the physical displacement of the tower locations, to a modal displacement $\mathbf{x}_\mathrm{m}\in \mathbb{R}^{N_\mathrm{m}}$. Here, $N_\mathrm{m}$ is the number of contributing modes considered.
%We adopt the convention that $z_1 = 1$. 
According to the order-reduction analysis~\cite{ruiterkampModellingControlLaterala}, $\mathbf{x}_{\mathrm{p}}$ can be expressed as:
%a product of the time-dependent $\mathbf{x}_\mathrm{m}$ and a term $\mathbf{S}$ that depends on the mode shape as:
\begin{equation}\label{eq:ModalProjection}
    \mathbf{x}_{\mathrm{p}} = \mathbf{S}^\top \mathbf{x}_\mathrm{m}\,,
\end{equation}
where $\mathbf{S} = [s_{il}] \in \mathbb{R}^{N_m\times N_\mathrm{l}}$ is a matrix in which each element $s_{il}$ represents the shape of the $i^\text{th}$ mode at height $z_l$. Such elements are defined as:
%\begin{equation}
%   s_{il} = \left[
%    \begin{array}{cccc}
%         \phi_{1,i}& \phi_{2,i} &  \dots &\phi_{N_d,i}
%    \end{array}
%    \right]
%    \left[
%    \begin{array}{c}
%         z_l^0 \\
%         z_l^1 \\
%         \vdots\\
%         z_l^{N_d-1}
%    \end{array}
%   \right]\,.
%\end{equation}
\begin{equation}
    s_{il} = \Phi_i^\top \left[ z_l^0 \quad  \hdots \quad z_l^{N_d-1} \right]^\top.
\end{equation}

% where $\Phi_i = [\phi_{1,i},\dots, \phi_{N_d,i}]^\top$ is the $i^\text{th}$ column of the mode-shape matrix $\mathbf{\Phi}\in\mathbb{R}^{N_\mathrm{d} \times N_\mathrm{m}}$, with $N_\mathrm{d}$ being the number of degrees of freedom.

The multi-mode tower vibration model  \cite{gawronskiAdvancedStructuralDynamics2004} can be then described compactly as:
\begin{equation}
\label{eq:ModalAnalysisDynamics}
\ddot{\mathbf{x}}_{\mathrm{m}} + \mathbf{M}_{\mathrm{m}}^{-1} \mathbf{D}_{\mathrm{m}} \dot{\mathbf{x}}_{\mathrm{m}} +\mathbf{ M}_{\mathrm{m}}^{-1} \mathbf{K}_{\mathrm{m}} \mathbf{x}_{\mathrm{m}}= \mathbf{B}_{\mathrm{m}} F_{\mathrm{T}}
\, ,
\end{equation}
where the diagonal matrices $\mathbf{M}_{\mathrm{m}}$, $\mathbf{K}_\mathrm{m}$, $\mathbf{D}_{\mathrm{m}}$ $ \in \mathbb{R}^{N_{\mathrm{m}} \times N_{\mathrm{m}}}$ are, respectively, the modal mass, stiffness and damping matrices.
Specifically, each diagonal element $m_{\mathrm{m},i}$ of $\mathbf{M}_\mathrm{m}$ is defined as~\cite{Zhang7117746,Branlard2019}:
\begin{equation}\label{eq:modMass}
    m_{\mathrm{m},i} = \sum_{l=1}^{N_{\mathrm{d}}} \rho(z_l)s_{il}^2 \Delta z_l H_{\mathrm{t}} \,,
\end{equation}
where $H_{\mathrm{t}}$ is the tower height; $\rho(z_l)$ is the tower mass density at height $z_l$,; and $\Delta z_l = z_l - z_{l-1}$ . 
The diagonal elements of $\mathbf{D}_{\mathrm{m}}$ and $\mathbf{K}_\mathrm{m}$ are assumed to be given.
The matrix $\mathbf{B}_{\mathrm{m}} \in \mathbb{R}^{N_{\mathrm{m}}}$ is the input matrix in modal analysis coordinate, defined as $\mathbf{B}_{\text{m}} = \mathbf{M}_{\text{m}}^{-1} \mathbf{\Phi}^\top \mathbf{B}_{\text{o}}$~\cite{gawronskiAdvancedStructuralDynamics2004}. 
The symbol $\mathbf{B}_{\text{o}} = [1 \quad \mathbf{0}^{1 \times N_{\mathrm{d}-1}}]^\top$ indicates that the aerodynamic thrust force $F_\mathrm{T}$ is applied at the top of the tower, as shown in Fig.~\ref{fig:TowerMAb}.
% can be seen as an input matrix to the displacement coordinates.
% As shown in Fig.\ref{fig:TowerMAb}, we only consider the thrust force to be acting on the top of the tower. 
% Thus, $\mathbf{B}_{\text{o}}=\left[1 \quad \mathbf{0}^{1 \times N_{\text{d}}-1} \right]^\top$.
Finally $F_\mathrm{T}$, which is the input to \eqref{eq:ModalAnalysisDynamics}, is calculated as:
\begin{equation}\label{eq:FT}
    F_\mathrm{T} = \frac{1}{2} \rho A C_\mathrm{t}(\lambda,\beta) v_\mathrm{w}^2\,,
\end{equation}
where $C_\mathrm{t}$ is the thrust coefficient, a nonlinear function of $\lambda$ and $\beta$, and is dependent on the physical characteristics of the turbine. \textcolor{zx}{Similar to $C_\mathrm{p}$, $C_\mathrm{t}$ can be derived and stored in a look-up table for control purposes.}

\section{EMPC Tower Damping Considering Multiple Vibration Modes}\label{sec:eMPC}
\noindent 
Having presented the WT model, let us now introduce the eMPC-based controller to simultaneously reduce fatigue loads on multiple tower locations.
Given the nonlinear nature of the WT dynamics defined in Section~\ref{sec:model}, we start by presenting a method to linearize them via a change of variables, introduced in \cite{Shaltout2017,Hovgaard2015};
following this, we mathematically formulate the objective function for tower load reduction;
finally, we define the full eMPC-based controller.
% \iffalse
%
% Having presented the dynamics of the wind turbine that are considered, and having summarized a change of variables that allows for their linear description, let us address the main focus of this paper, an eMPC framework capable of reducing the loading of multiple locations of the turbine tower.
% In the remainder of this section, we start by mathematically formalizing the objective, after which we describe in detail the eMPC-based controller.
% \fi
% In  this  section,  the eMPC framework is presented for fatigue loads reduction on specific locations. The basic assumptions are listed here:
% \begin{assumption}
%     There are displacement and velocity sensors available on the specific tower locations.
%     \label{Assumption1}
% \end{assumption}
% \begin{assumption}
%     The tower natural frequencies, corresponding mode shape and damping ratios are known.
%     \label{Assumption2}
% \end{assumption}
 %\begin{itemize}
 %The proposed control scheme is shown in Fig.\ref{Framework}. %The measurements of $P_\mathrm{g}, \omega_\mathrm{g}$ and $\mathbf{x}_\mathrm{m}, \mathbf{v}_\mathrm{m}$ are fed to the eMPC algorithm at each time step, while $P_{\mathrm{r}}$ is approximated by (\ref{Eq:Pr}) and then is fed to the eMPC.
 
    %\begin{figure}[t]
    % \centering
    %\includegraphics[width=\columnwidth]{figs/Framework.pdf}
      
    %  \caption{Block diagram of the eMPC-based controller.}
    %  \label{Framework}
   %\end{figure}

\subsection{Convex Constraints Formulation}\label{sec:eMPC:cvx}
\noindent To simplify the analysis of the eMPC-based controller, the nonlinear model of the WT can be made linear by introducing the \textit{kinetic energy}, while keeping the constraints convex~\cite{Shaltout2017},~\cite{Hovgaard2015}. 
%Such a transformation has been used for power maximization in~\cite{Hovgaard2015}, and exploited in~\cite{Shaltout2017} for power maximization with tower-bottom fatigue loads mitigation.
%We express the drive train model of the WT in terms of the power flows and energy exchanges between the wind and the wind turbine~\cite{Shaltout2017}.
%To achieve this, we introduce 
The kinetic energy $K$ stored in the generator is defined as:
\begin{equation}
    K = \frac{J}{2}\omega_{\mathrm{g}}^2
    \, .
    \label{eq:Kenergy}
\end{equation}
By substituting~\eqref{eq:Kenergy} into the drive train dynamics defined in~\eqref{DriveTrainDynamics}, dynamics of $K$ is derived as:
\begin{equation}
    \label{NewDT}
\dot{K}=J \omega_{\mathrm{g}} \dot{\omega}_{\mathrm{g}}=\omega_{\mathrm{g}}\left(\frac{1}{G} T_{\mathrm{r}}-T_{\mathrm{g}}\right)=P_{\mathrm{r}}-\frac{1}{\eta_{\mathrm{g}}} P_{\mathrm{g}}
\,.
\end{equation}
Here, the introduction of $P_\mathrm{r}$ and $P_\mathrm{g}$, considered as directly controllable inputs, allows for a linear formulation of the dynamics.
%Note that this is an approximation, as $P_\mathrm{r}$ and $P_\mathrm{g}$ are the result of the control action on $\beta$ and $T_\mathrm{g}$, as discussed previously. 
%However, once desired values for $P_\mathrm{r}$ and $P_\mathrm{g}$ are computed by the controller, inverse mappings from $P_\mathrm{r}$ and $P_\mathrm{g}$ must be used to find $\beta$ and %$T_\mathrm{g}$, so long as $\omega_\mathrm{g}$ and $v_\mathrm{w}$ are known.
%These inverse mappings can be derived from \eqref{Eq:Pr} and \eqref{eq:Pg} together with the definition of $K$ in~\eqref{eq:Kenergy}.

Finally, to complete the linearization of the dynamics, we must ensure that the modal dynamics in \eqref{eq:ModalAnalysisDynamics} are linear with respect to the kinetic energy~\cite{Shaltout2017}.
To do this, we define the following approximation of the thrust force:
\begin{equation}\label{eq:FtHat}
    \hat{F}_\mathrm{T} = \zeta_1 P_\mathrm{r} + \zeta_2 K + \zeta_3 \,,
\end{equation}
where $\zeta_1 \, , \zeta_2 \, , \zeta_3 \in \mathbb{R}$ are derived from the linearization of \eqref{eq:FT} around an operating point. 
By defining $\mathbf{v}_\mathrm{m} = \dot{\mathbf{x}}_\mathrm{m}$, we have:
\begin{equation}\label{eq:vM}
    \dot{\mathbf{v}}_\mathrm{m} = 
    \mathbf{M}_{\mathrm{m}}^{-1} \left[\mathbf{\Phi}^\top \left( \zeta_1 P_{\mathrm{r}}+\zeta_2 K+\zeta_3 \right) - \mathbf{D}_{\mathrm{m}} \mathbf{v}_{\mathrm{m}} - \mathbf{K}_{\mathrm{m}} \mathbf{x}_{\mathrm{m}}  \right]
    \,.
\end{equation}
Hence, a state space representation of the turbine dynamics can be defined as:
\begin{equation}\label{eq:dyn}
    \dot{\mathbf{x}}(t) = A\mathbf{x} + B \mathbf{u}(t)\,,
\end{equation}
where $\mathbf{x}= [K,\mathbf{x}_m^\top, \mathbf{v}_m^\top]^\top \in \mathbb{R}^{2N_m+1}$, $\mathbf{u} = [P_\mathrm{r}, P_\mathrm{g}]^\top \in \mathbb{R}^2$, and $A$ and $B$ are derived from \eqref{NewDT} and \eqref{eq:vM}.
%\begin{assumption}\label{ass:know}
%    All parameters related to the wind turbine are known.
    % The tower natural frequencies, mode shape and damping factors are known exactly.
%    $\hfill\triangleleft$
%end{assumption}

We now show that the constraints \eqref{eq:cstr:state}-\eqref{eq:cstr:inputBeta} remain convex.
Given the definition of $K$ in \eqref{eq:Kenergy}, the state constraints expressed in \eqref{eq:cstr:state} can be  rewritten as:
\begin{equation}\label{eq:cstr:K}
    \frac{J}{2} \omega_{\mathrm{g}, \min }^{2} \leq K \leq \frac{J}{2} \omega_{\mathrm{g}, \max }^{2}  \,,
%\label{KConstraints}
\end{equation}
%with $\omega_\mathrm{g,max}>\omega_\mathrm{g,min}\geq 0$, given that by %construction the turbine blades always rotate in the same direction.
Moreover, the input constraints \eqref{eq:cstr:inputTg}-\eqref{eq:cstr:inputBeta} can be rewritten for $P_\mathrm{r}$ and $P_\mathrm{g}$ as:
\begin{equation}\label{eq:cstr:PrPg}
    \begin{cases}
        0\leq P_\mathrm{r} \leq \hat{P}_{\mathrm{av}}(v_\mathrm{w},K)\\
        0\leq P_\mathrm{g} \leq \eta_\mathrm{g} T_{\mathrm{g,max}}\sqrt{\frac{2}{J}K}
    \end{cases}\,,
\end{equation}
where $\hat{P}_\mathrm{av}(v_\mathrm{w},K)$ is a convex approximation of  $P_\mathrm{av}(v_\mathrm{w},K)$, the available wind power.
Similarly, the physical bound of $\hat{F}_\mathrm{T}$ as defined in \eqref{eq:FtHat} must satisfy:
\begin{equation}\label{eq:cstr:FT}
    0 \leq \hat{F}_T \leq F_{T,\mathrm{max}} \,,
\end{equation}
where $F_{T,\mathrm{max}}$ is the maximum thrust force, given $v_\mathrm{w}$ and $K$, as defined in~\cite{Shaltout2017}, i.e.
\begin{equation}
\label{FTMax}
{F}_{T, \max } =\max _{\beta_{\min } \leq \beta \leq \beta_{\max }} 0.5 \rho A C_\mathrm{t}(K, \beta) v_{\mathrm{w}}^{2}
\, .
\end{equation}

In conclusion, a convex constraint function $C(\mathbf{x},\mathbf{u})$, to be used in eMPC, can be constructed such that $C(\mathbf{x},\mathbf{u})\leq 0$ ensures that \eqref{eq:cstr:K}, \eqref{eq:cstr:PrPg} and \eqref{eq:cstr:FT} hold.

\iffalse
Thus, to summarize, it is possible to rewrite the dynamics of WT's drive train and its tower vibration as
\begin{equation}
\label{LinearModel}
\left\{\begin{array}{l}
\dot{K}=P_{\mathrm{\mathrm{r}}}-\frac{1}{\eta_{\mathrm{g}}} P_{\mathrm{g}} \\
\dot{\mathbf{x}}_{\mathrm{m}}=\mathbf{v}_{\mathrm{m}} \\
\dot{\mathbf{v}}_{\mathrm{m}}  = \mathbf{M}_{\mathrm{m}}^{-1} \left[\mathbf{\Phi}^\top \left( q P_{\mathrm{r}}+r K+s \right) - \mathbf{D}_{\mathrm{m}} \mathbf{v}_{\mathrm{m}} - \mathbf{K}_{\mathrm{m}} \mathbf{x}_{\mathrm{m}}  \right]
\end{array}\right.
\, .
\end{equation}
where $\mathbf{v}_\mathrm{m}$ is the modal velocity.

Having presented a method, introduced in \cite{Hovgaard2015}, to redefine the WT and modal dynamics such that they are linear, and reformulated the constraints such that they are convex, we can now introduce the eMPC-based controller capable of maximizing power generation while limiting the loads at multiple locations of the turbine tower.
\fi

\subsection{Load reduction at multiple tower locations}
\noindent 
The objective of our proposed controller is to reduce the fatigue loads at multiple locations. Let us start by introducing the \emph{Tower Fore-Aft Moment} (TFAM) at location $z_l$ of the tower.
The cyclic loads, \textit{i.e.}, the variation of $\mathrm{TFAM}(z_l)$ over time~\cite{SchlipfLIDAR2013} and at multiple heights $z_l, l \in \{1,\dots,N_l\}$ are taken into account as they are related to fatigue accumulation:
\begin{equation}
    \label{eq:TFAM}
    \begin{split}
        \frac{d}{dt}\mathrm{TFAM}(z_l) = (H_{\mathrm{t}}-z_l)(d_l (\ddot{x}_{\mathrm{p},\mathrm{top}}-\ddot{x}_{\mathrm{p},l}) \\
        + k_l  (\dot{x}_{\mathrm{p},\mathrm{top}}-\dot{x}_{\mathrm{p},l})) \,,
    \end{split}
\end{equation}
where $x_{\mathrm{p},\mathrm{top}} = H_{\mathrm{t}}$, and $d_l$, $k_l$ are the damping and stiffness coefficients, which are constant for $z_l$. Such loads can be effectively reduced by minimizing the term $\dot{x}_{\mathrm{p},\mathrm{top}}-\dot{x}_{\mathrm{p},l}$~\cite{Shaltout2017}. Let $\mathbf{v}_{\mathrm{p}} = \dot{\mathbf{x}}_{\mathrm{p}}$. Since the loads at the tower base are the highest, $v_{\mathrm{p},\mathrm{bottom}} = 0$ must be included in $\mathbf{v}_{\mathrm{p}}$.
As a proxy for minimizing the TFAM, we will here minimize $\mathbf{v}_{\mathrm{p}}$, by assigning a larger weight for minimizing $v_{\mathrm{p}}$ (details are explained in Section~\ref{sec:simulation}-B).
With \eqref{eq:ModalProjection} in mind, we can thus define the following objective
% \begin{equation}
%     O_v = \mathbf{w}^\top \mathbf{v}_{\mathrm{m}}^\top \mathbf{S} \mathbf{S}^\top \mathbf{v}_{\mathrm{m}}
% \end{equation}
\begin{equation}
    O_v = \textcolor{ajg}{\mathbf{v}_\mathrm{p}^\top \mathbf{W} \mathbf{v}_\mathrm{p} =}
    \mathbf{v}_{\mathrm{m}}^\top \mathbf{S} \textcolor{ajg}{\mathbf{W}} \mathbf{S}^\top \mathbf{v}_{\mathrm{m}}
\end{equation}
where $\mathbf{W}=\mathrm{diag}_{i \in \{1. \dots, N_l\}} [w_i]$ are weights allowing to account for the relative importance of multiple tower locations.

\subsection{Load-limiting eMPC}
\noindent 
We are now ready to present the eMPC-based controller to achieve power maximization with tower load limiting at multiple locations.

As discussed in Section~\ref{sec:eMPC:cvx}, the WT dynamics can be described as the linear system in \eqref{eq:dyn}, and a convex constraint $C(\mathbf{x},\mathbf{u})\leq 0$ can be defined.
In addition, we include in $C(\mathbf{x},\mathbf{u})$ the following
\begin{equation}
    0 \leq K  \leq \frac{J}{2}\omega_\mathrm{g,rated}^2 + \epsilon\,,
\end{equation}
where $\epsilon$ is a variable in the objective function, used to limit the turbine's oversping, as is further detailed in the following.

Finally, before moving on to the definition of the eMPC-based controller, let us define its objective function.
We consider a scalar weighted sum of multiple objectives, where the weights are defined such that appropriate tradeoffs between conflicting goals can be achieved.
Specifically, we define 
\begin{equation}
    \begin{split}
        O(\mathbf{x},\mathbf{u}) \doteq &\alpha_1 P_\mathrm{g}
        + \alpha_2 \hat{P}_\mathrm{av}(v_\mathrm{w},K) - \alpha_3 \dot{P}_\mathrm{g}^2
        - \alpha_4 \dot{P}_\mathrm{r}^2\\
        &- \alpha_5 \epsilon - O_v(\mathbf{x})\,,
    \end{split}
\end{equation}
where $\alpha_{\gamma}, \gamma \in\{1,\dots,5\}$ are appropriately defined weights, to be tuned together with $w_l, l \in\{1,\dots,N_l\}$. 
When maximizing the objective function $O(\mathbf{x},\mathbf{u})$, the first terms in $P_\mathrm{g}$ and $\hat{P}_\mathrm{av}$ determines the maximization of the power output of the turbine. Next two terms penalize their rate of change, while the fifth one penalizes the turbine's deviation from its rated rotational speed. Finally, the last term minimizes the velocity of the displacement at $N_l$ tower locations, and thus the TFAM.

In line with standard formulation of model predictive control, the control law is formulated as a \emph{Finite-Horizon Optimal Control Problem} (FHOCP):
\begin{subequations}\label{eq:FHOCP}
    \begin{align}
        \displaystyle\max_{\mathbf{U},\epsilon}\quad &\displaystyle\sum_{q = 0}^{N_{\mathrm{p}}-1} O(\bar{\mathbf{x}}(q),\mathbf{u}(q)) \,, \\
        \text{s.t.} \quad&\bar{\mathbf{x}}(q+1) = A_d\bar{\mathbf{x}}(q) + B_d \mathbf{u}(q) \,, \label{eq:FHOCP:dyn}\\
        &C(\bar{\mathbf{x}}(q), \mathbf{u}(q))\leq 0 \,, \forall q \in \{0,\dots,N_{\mathrm{p}}-1\} \,, \label{eq:FHOCP:cstr}\\
        &\bar{\mathbf{x}}(0) = \mathbf{x}(t_k) \,, \label{eq:FHOCP:init}\\
        &\bar{\mathbf{x}}(N_{\mathrm{p}}) = \mathbf{x}_s\,, \label{eq:FHOCP:fin}
    \end{align}
\end{subequations}
where $N_\mathrm{p}$ is the finite horizon,  $\bar{\mathbf{x}}(q) = [\bar{K}(q),\bar{\mathbf{x}}_\mathrm{m}^\top(q),\bar{\mathbf{v}}_\mathrm{m}^\top(q)]^\top,\, q \in \{0,\dots, N_{\mathrm{p}}\}$ is the predicted state of the system at time instant $q$, 
$(A_d,B_d)$ are the result of discretization of \eqref{eq:dyn} with a sampling time $T_s$. 
$\mathbf{x}(t_k)$ is the state at time $t_k$, which is reconstructed from the measurements $\omega_{\mathrm{g}}(t_k)$, $\mathbf{x}_{\mathrm{p}}$, $\mathbf{v}_{\mathrm{p}}$ based on \eqref{eq:Kenergy} and \eqref{eq:ModalProjection}.
and $\mathbf{U} = [\mathbf{u}^\top(0),\dots \mathbf{u}^\top(N_{\mathrm{p}}-1)]^\top \in \mathbb{R}^{2N_{\mathrm{p}}}$.

The constraints of the FHOCP are defined as follows:
\eqref{eq:FHOCP:dyn} represents the dynamics with which the state is predicted,
\eqref{eq:FHOCP:cstr} guarantees that the constraints are satisfied,
\eqref{eq:FHOCP:init} sets the initial state of the nominal model to be the same as the measured state at time $t_k$. 
\eqref{eq:FHOCP:fin} is a terminal constraint on the predicted state, which ensures that the terminal predicted state is the same as $\mathbf{x}_s$, the optimal steady-state state. This is calculated by solving:
\begin{equation}\label{eq:Opt_steadyState}
    \begin{array}{cl}
         \displaystyle
         \max_{\mathbf{x}_s,\mathbf{u}_s} &O(\mathbf{x}_s,\mathbf{u}_s)\\
         \text{s.t.} & 
         \mathbf{x}_s = A_d \mathbf{x}_s + B_d\mathbf{u}_s\\
         &C(\mathbf{x}_s,\mathbf{u}_s)\leq 0
    \end{array}\,.
\end{equation}

The terminal constraint \eqref{eq:FHOCP:fin} is included in the FHOCP, to ensure that the optimal trajectories over the prediction horizon do not exhibit \textit{turnpike} behavior, and ensures recursive feasibility of the FHOCP \eqref{eq:FHOCP}.

The solution to \eqref{eq:FHOCP} is computed at every time step $t_k$, the optimal value $\mathbf{U}^*$ is found, and $\mathbf{u}(t_k) = \mathbf{u}^*(0)$ is applied to the system.
For each of the following time instants, the problem \eqref{eq:FHOCP} will roll ahead and the procedure is repeated at the next time instant.
In the proposed scheme the model was transformed resulting in linear dynamics, and convex constraints and a concave objective function are derived. Therefore, the problem can be solved globally and in a computationally efficient way~\cite{boyd2004convex}.

\textcolor{akp}{
Since the dynamics and constraints of the FHOCP are formulated in power and energy variables as detailed in Section~\ref{sec:eMPC:cvx}, it is necessary for $\mathbf{u}(t_k)$ to be translated back into the original, usable WT control signals, namely $T_\mathrm{g}^* = P_\mathrm{g}/(\eta_\mathrm{g} \sqrt{2K^*/J})$\footnote{With a slight abuse of notation, $K^*$ here represents the one-step ahead prediction of $K$.} and $\beta^*=\Psi(P_\mathrm{r}^*,v_\mathrm{w},K^*)$.
The pitch look-up table $\Psi$ contains the inverse nonlinear mapping involving $C_\mathrm{p}$, only defined for the pitch range \eqref{eq:cstr:inputBeta}.
Thus, it is ensured that $\beta$ will not violate its operational bounds.}

\section{Simulation Results and Analysis}\label{sec:simulation}
\noindent In this section, the performance of the proposed eMPC approach is demonstrated on a simplified NREL's 5MW WT dynamic model.
The eMPC-based controller proposed in Section~\ref{sec:eMPC} with a two-mode model of the tower vibration dynamics is compared to two eMPC-based controllers in literature: one with single mode damping, proposed in~\cite{Shaltout2017}, and one without tower damping, designed in~\cite{Hovgaard2015}.
% The case studies of (1) two modes damping control, (2) one mode damping control~\cite{Shaltout2017} and (3) no tower damping control activated~\cite{Hovgaard2015} are implemented for comparisons.
%The controllers are implemented under the same wind conditions. 

\subsection{Turbine Configuration}
%\noindent The tower model parameters of the NREL's 5MW wind turbine model are listed in Table.\ref{ModelParameter}. 
\noindent \textcolor{zx}{The natural frequency of the first and second mode are 0.3240 Hz and 2.9003 Hz, respectively. }\textcolor{zx}{Other structural and dynamic model parameters, such as the tower mass and damping ratio, can be found in~\cite{jonkmanDefinition5MWReference2009}.}

Without loss of generality, two tower measurements considered in the case study to illustrate the load mitigation performance: i. the tower top ($z_1 = 1$) where the dominant displacement occur, and ii. the location ($z_2 = 0.72$) where the maximum displacement of the second mode shape is observed. 
In practice, other tower locations where the measurement device is instrumented can be selected for load reduction, depending on the user's need.
% The first one is the tower top location, which is $z_1 = H_{\mathrm{t}}$. 
% The second one is a location near the maximum displacement of the second mode shape, that is, $z_2 = 0.72H_{\mathrm{t}}$.
% density in Table~\ref{TowerProperties} is used to derive the modal mass $\mathbf{M}_\mathrm{m}$ through~\eqref{eq:modMass}.
% Then the density in Table~\ref{TowerProperties} and $s_i(z)$ are utilized to derive the modal mass $M_{m}(i,i)$ for the $i^{\mathrm{th}}$ mode~\cite{Zhang7117746}~\cite{Branlard2019}:

%begin{table}[t]
%\caption{NREL's 5 MW wind turbine model parameters~\cite{jonkmanDefinition5MWReference2009}}
%\label{ModelParameter}
%\begin{center}
%\begin{tabular}{|c||c|}
%\hline
%Parameter & Value \\
%\hline
%1st tower fore-aft natural frequency, $f_{1}$ & 0.3240 Hz\\
%2nd tower fore-aft natural frequency, $f_2$ & 2.9003 Hz\\
%1st mode damping ratio, $d_1$ & 1\% \\
%2nd mode damping ratio, $d_2$ & 1\% \\
%\hline
%\end{tabular}
%\end{center}
%\end{table}
 
% \begin{equation}
%     M_{\mathrm{m}}(i,i) = \int_{0}^{H_{\mathrm{t}}} \rho(z) s_i(z)^{2} d z
%     \, ,
% \end{equation}
% where $\rho(z)$ is the tower mass density.
\subsection{Simulation Configuration}
\noindent A uniform wind flow, whose velocity follows a staircase profile, is considered. 
The stepwise wind speed increases from $6$ to $17~\mathrm{m/s}$, with each step lasting for 100 s. 
The weights $\alpha_{\gamma} \, (\gamma=\{1,...,5\})$ and $w_l, l = \{1,2\}$ are tuned to find a suitable trade-off between power regulation and tower loads reduction. 
For $\alpha_{\gamma}$, the weights are selected to guarantee each term with the same order of magnitude. 
While designing $w_l$, we focus more on the tower bottom load-reduction because it is the location where the strongest fatigue loads are suffered. 
Thus, we choose $w_1 - w_2 >2w_2$ to guarantee the proxy mentioned in Section~\ref{sec:model}-B. 
Besides, $w_l$ can not be too large because we do not want to sacrifice too much generated power.
Based on the rules explained above, we choose the weights presented in Table \ref{tb:Weights}.
%\begin{figure}[htpb]
%      \centering
      %\includegraphics[width=\columnwidth]{figs/Location1.pdf}
%      \includegraphics[width=\columnwidth]{figs/WindSpeed.pdf}
%      \caption{Wind speed profile.}
%      \label{fig:Vw}
%   \end{figure}

\begin{figure}[tbph]
    \centering
    \subfigure[$P_{\text{g}}$ comparison.]{
    \begin{minipage}[t]{0.45\columnwidth}
    %\begin{minipage}[t]{0.45\textwidth}
    \centering
    \includegraphics[width=\linewidth]{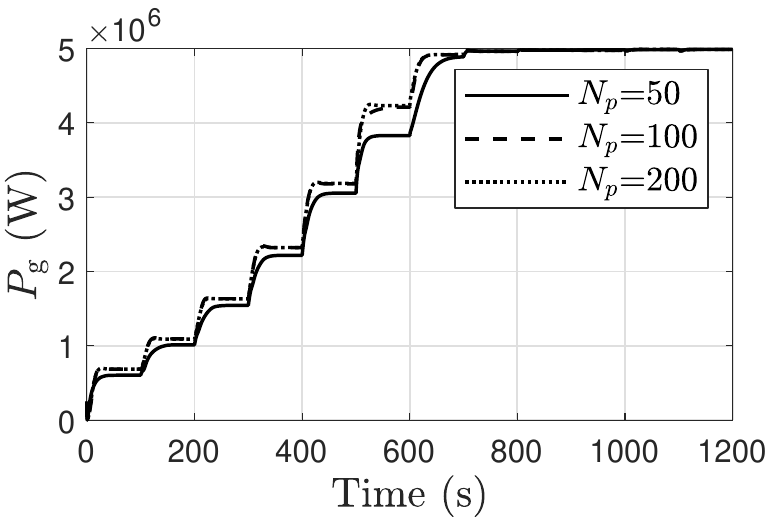}
    \end{minipage}
   }  
   \hfill
   \subfigure[Computation time comparison.]{
    \begin{minipage}[t]{0.45\columnwidth}
    %\begin{minipage}[t]{0.45\textwidth}
    \centering
    \includegraphics[width=\linewidth]{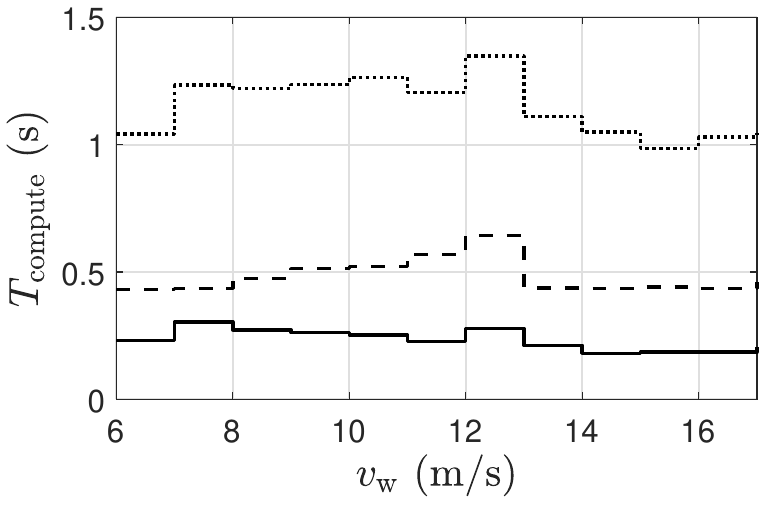}
    \end{minipage}
   }  
    \caption{Comparison of $N_p=50$, $N_p=100$, $N_p=200$ .}
    \label{fig:NpComparison}
\end{figure}

The simulation platform is MATLAB/Simulink software. The computer configuration is a laptop with a CPU i7-8665U, 2.11 GHz frequency and a 8 GB memory capacity. The sampling time of the simulation is $0.2~\mathrm{s}$. Regarding the definition of the prediction horizon, \textcolor{ajg}{we remark that this should be chosen as a compromise between the computational time required to solve \eqref{eq:FHOCP} and system performance.
To demonstrate this trade-off,} we compare three different values of $N_{\mathrm{p}}$: a. $N_{\mathrm{p}}=50$, b. $N_{\mathrm{p}}=100$, c. $N_{\mathrm{p}}=200$. 
The results of this comparison are shown in Fig.~\ref{fig:NpComparison}, from which it is found that the controller with $N_{\mathrm{p}}=100$ shows similar performance to the one with $N_{\mathrm{p}}=200$, but at lower computation cost.
Therefore, $N_{\mathrm{p}} = 100$ is selected for the comparison study, implying a $20\mathrm{s}$ prediction horizon.

\begin{table}[tbhp]
\caption{Weights in objective function}
\label{tb:Weights}
\centering
\begin{tabular}{ccccccc}
\hline
 $\alpha_1$ & $\alpha_2$ & $\alpha_3$ & $\alpha_4$  & $\alpha_5$ & $w_1$%$\alpha_6$ 
 & $w_2$ %$\alpha_7$ 
 \\ \hline
 1 & 1 & 1 & 0.01 & 100 & 100 & 20   \\\hline
\end{tabular}
\end{table}

\subsection{Controller Performance}
\noindent 
\begin{figure}[tbph]
      \centering
      \includegraphics[width=0.75\columnwidth]{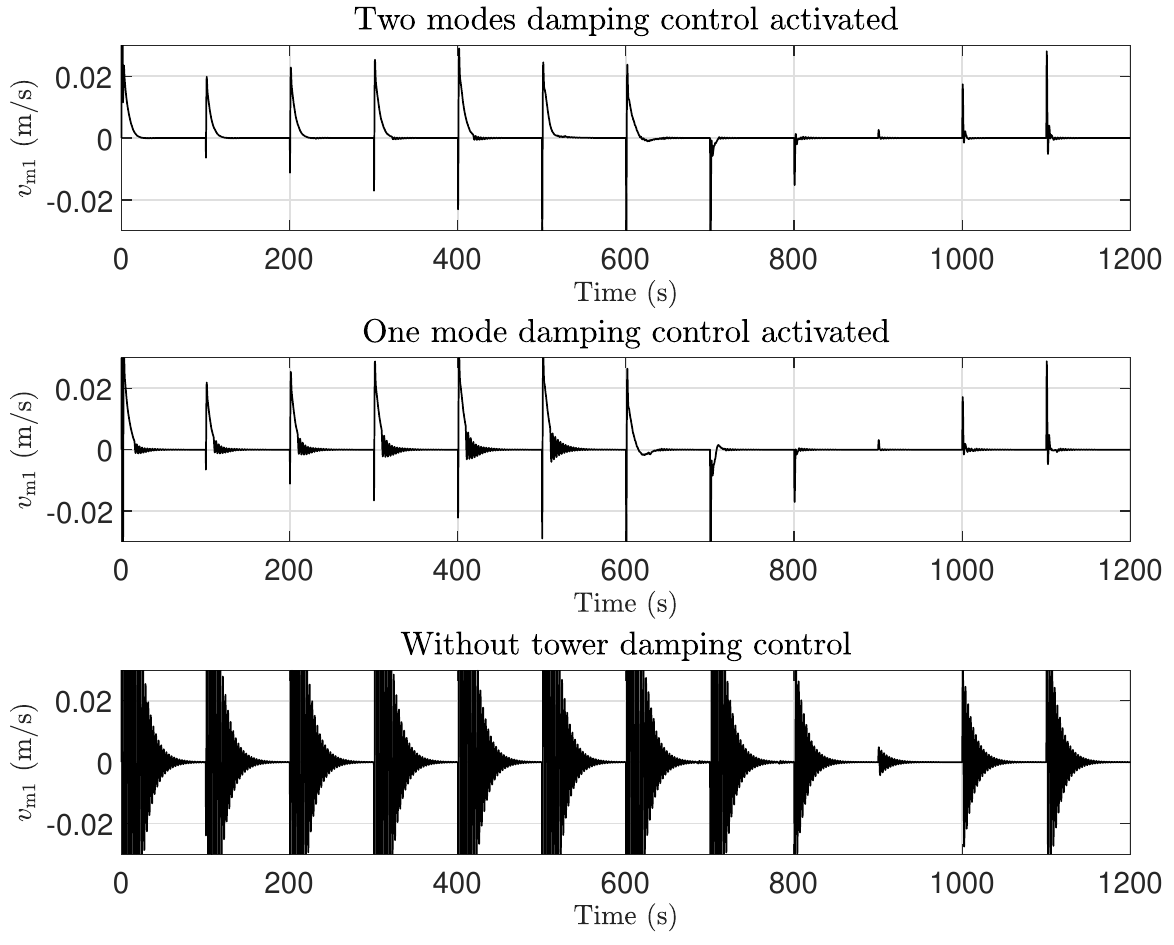}
      \caption{Comparison of $v_1$ on location $z_1=1$.}
      \label{Comparison1}
   \end{figure}
\begin{figure}[tbph]
      \centering
      \includegraphics[width=0.75\columnwidth]{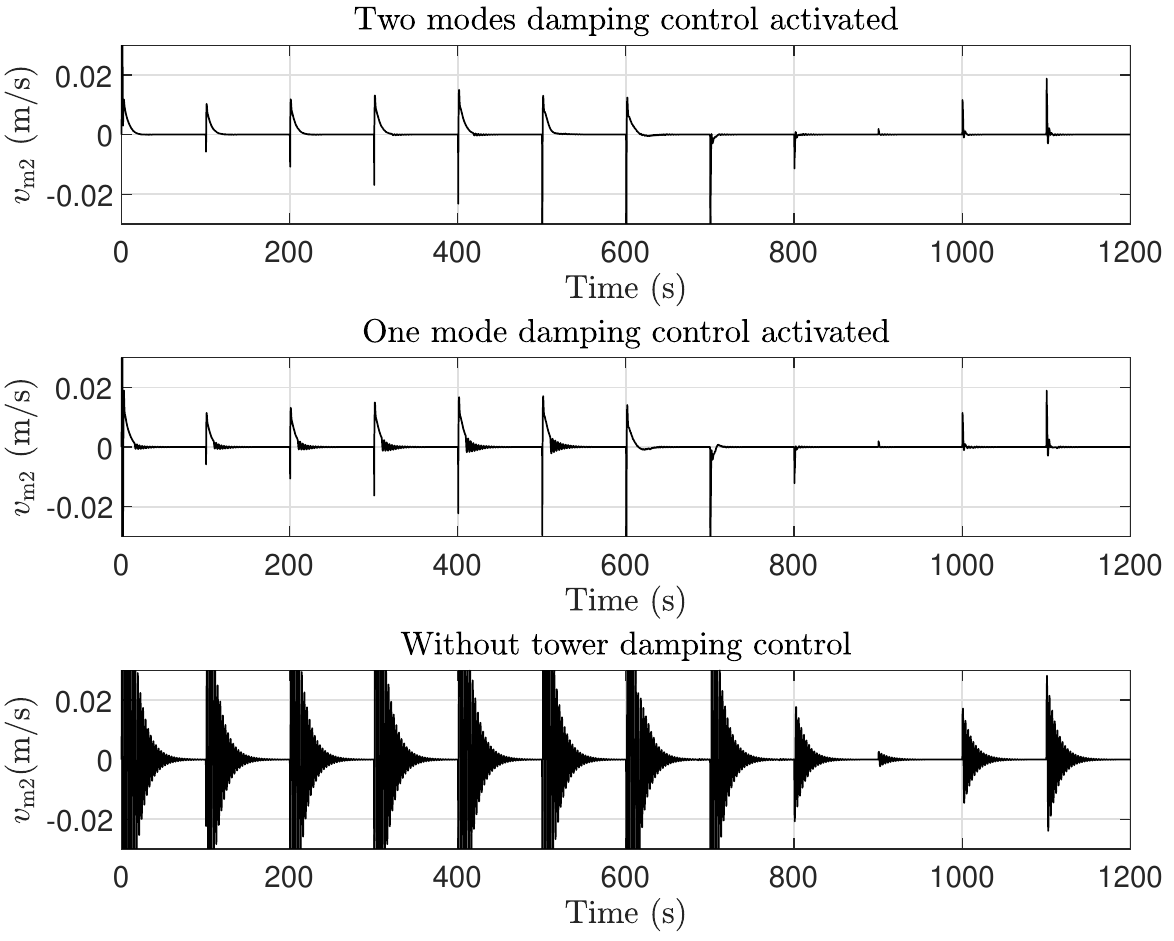}
      \caption{Comparison of $v_2$ on location $z_2=0.72H_\mathrm{t}$.}
      \label{Comparison2}
   \end{figure}
\begin{figure}[tbph]
  \centering
  \subfigure[$v_{\mathrm{w}} = 7$ m/s]{
  \begin{minipage}[t]{0.42\columnwidth}
  %\begin{minipage}[t]{0.45\textwidth}
  \centering
  \includegraphics[width=\linewidth]{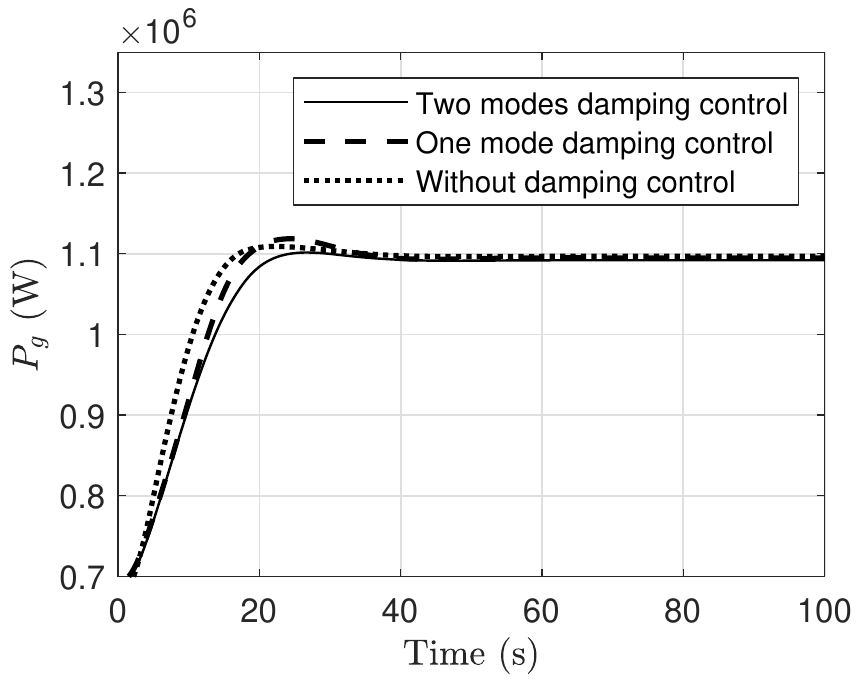}
  \label{ComparisonPga}
  \end{minipage}
  }
  \hfill
  \subfigure[$v_{\mathrm{w}} = 10$ m/s]{
  \begin{minipage}[t]{0.42\columnwidth}
  %\begin{minipage}[t]{0.45\textwidth}
  \centering
  \includegraphics[width=\linewidth]{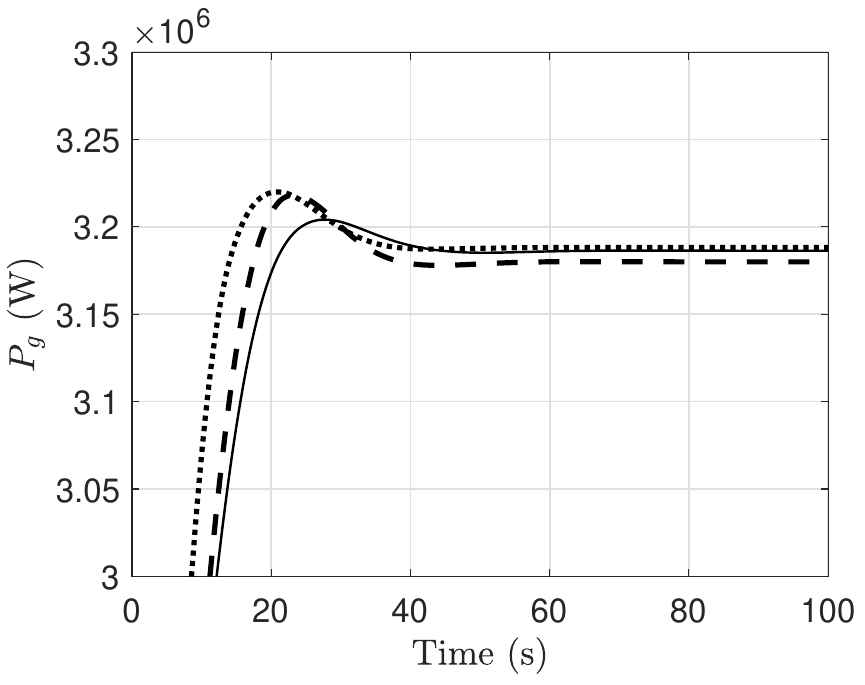}
  \label{ComparisonPgb}
  \end{minipage}
  } 
  \vfill
    \subfigure[ $v_{\mathrm{w}} = 14$ m/s]{
  \begin{minipage}[t]{0.42\columnwidth}
  %\begin{minipage}[t]{0.45\textwidth}
  \centering
  \includegraphics[width=\linewidth]{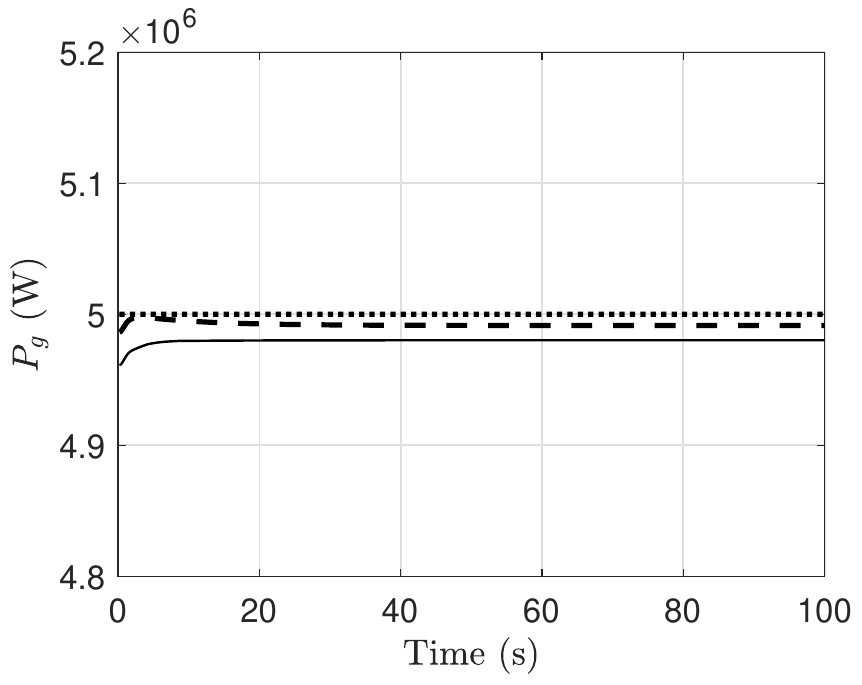}
  \label{ComparisonPgc}
  \end{minipage}
  }
  \hfill
  \subfigure[$v_{\mathrm{w}} = 17$ m/s]{
  \begin{minipage}[t]{0.42\columnwidth}
  %\begin{minipage}[t]{0.45\textwidth}
  \centering
  \includegraphics[width=\linewidth]{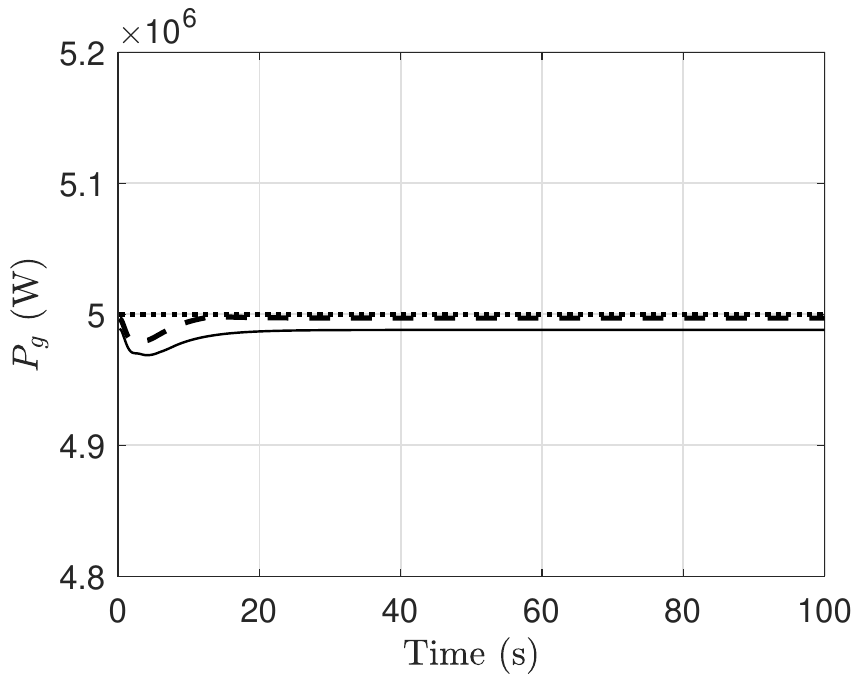}
  \label{ComparisonPgd}
  \end{minipage}
  }  
  \caption{Comparison of $P_g$ at different wind speeds.}
  \label{ComparisonPg}
%\end{subfigure}
\end{figure}

First, the trajectories of the velocities and accelerations are presented in Figs.~\ref{Comparison1} and~\ref{Comparison2}. It shows that the structural load at both locations is alleviated by the two-modes-damping control strategy: indeed, the amplitude and oscillation of the velocities of both locations are significantly reduced. 
With~\eqref{eq:TFAM} in mind, this leads to the fatigue loads at both locations being alleviated.
Furthermore, Fig.~\ref{ComparisonPg} illustrates the performance of the power generation. 
In Figs.~\ref{ComparisonPga} and~\ref{ComparisonPgb}, at $v_{\mathrm{w}} = 7$ m/s and $10$ m/s, $P_\mathrm{g}$ at steady states stays the same. 
In Figs.~\ref{ComparisonPgc} and~\ref{ComparisonPgd}, at $v_{\mathrm{w}} = 14$ m/s and $17$ m/s, there is a minor power loss around 0.5$\%$, which can be considered as negligible impacts on the power generation.
The power generation at other wind speeds are similar, which is omitted for brevity. 
In summary, a small amount of $P_{\text{g}}$ is sacrificed for tower load-reduction at above-rated wind speeds, while there is no power loss introduced by the proposed load-limiting control at below-rated wind speeds.
%\begin{figure}
%    \centering
%    \subfigure[Comparison of $v_1$ on location $z_1=H_t$.]{
%    \centering
%    \label{fig:Comparison1}
 %   \includegraphics[width=0.75\columnwidth]{figs/Location1.pdf}}
%    \vfill
%    \subfigure[Comparison of $v_2$ on location $z_2=0.72H_t$.]{
 %   \centering
 %   \label{fig:Comparison1}
  %  \includegraphics[width=0.75\columnwidth]{figs/Location2.pdf}}
  %  \caption{Comparison of velocity on the two locations.}
 %   \label{fig:Comparison}
%\end{figure}
Therefore, it is concluded that the proposed eMPC-based controller performs efficient load reduction at multiple tower locations, without significant effects on the power regulation.

\begin{figure}[tphb]
    \centering
    \subfigure[$K(t)$ comparison.]{
    \begin{minipage}[t]{0.45\columnwidth}
    %\begin{minipage}[t]{0.45\textwidth}
    \centering
    \includegraphics[width=\linewidth]{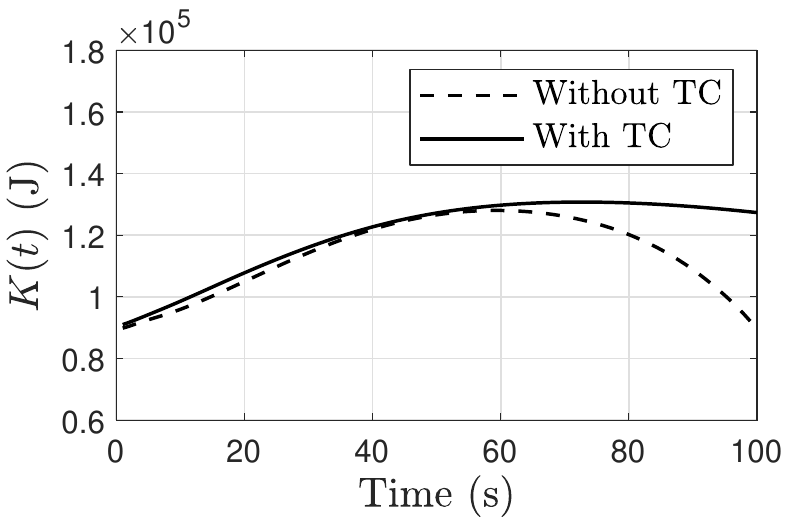}
    \label{fig:TCcomparisona}
    \end{minipage}
   }  
   \hfill
   \subfigure[$v_{\mathrm{m}1}$ comparison.]{
    \begin{minipage}[t]{0.45\columnwidth}
    %\begin{minipage}[t]{0.45\textwidth}
    \centering
    \includegraphics[width=\linewidth]{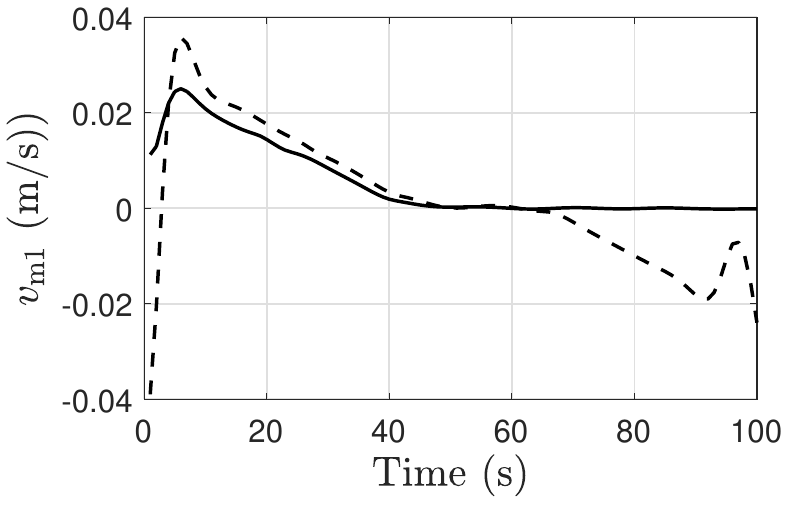}
    \label{fig:TCcomparisonb}
    \end{minipage}
   } 
   \vfill
    \centering
    \subfigure[$P_{\mathrm{g}}$ at $v_{\mathrm{w}}=16$ m/s.]{
    \begin{minipage}[t]{0.45\columnwidth}
    %\begin{minipage}[t]{0.45\textwidth}
    \centering
    \includegraphics[width=\linewidth]{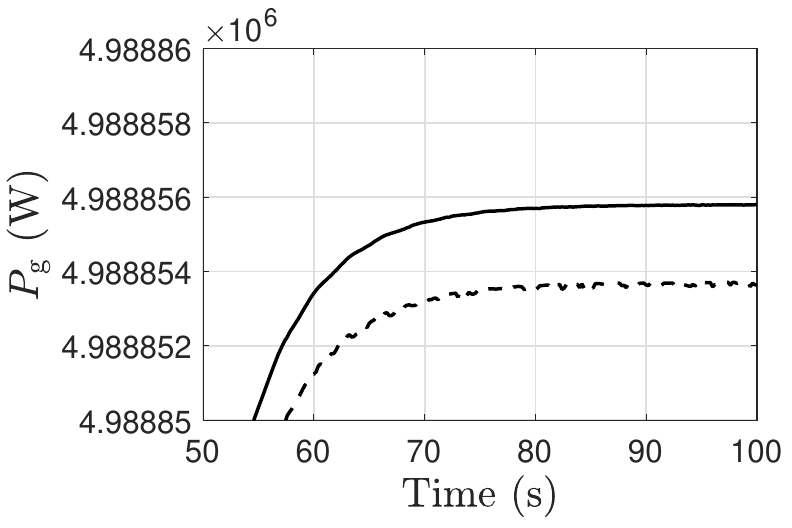}
    \label{fig:TCcomparisonc}
    \end{minipage}
   }  
   \hfill
   \subfigure[Computation time comparison.]{
    \begin{minipage}[t]{0.45\columnwidth}
    %\begin{minipage}[t]{0.45\textwidth}
    \centering
    \includegraphics[width=\linewidth]{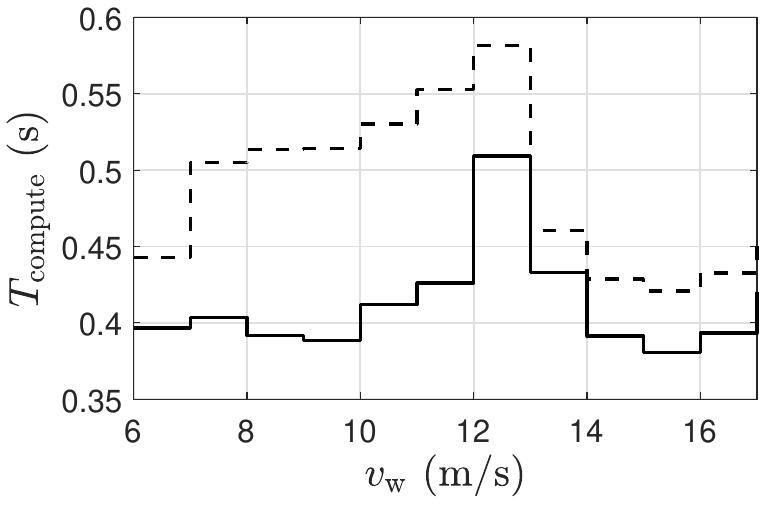}
    \label{fig:TCcomparisond}
    \end{minipage}
   } 
    \caption{Comparison of performance between the eMPC-based controllers with and without terminal constraints}
    \label{fig:TCcomparison}
\end{figure}
%\subsection{Turbulent Wind Condition}
%To show more insights, we simulate the proposed eMPC controller at turbulent %wind speeds.
\subsection{The Importance of Terminal Constraints}
\noindent As discussed in Section~\ref{sec:eMPC}, %without terminal constraints, the definition of the eMPC objective function may lead %to turnpike behavior over the prediction horizon. %Thus, 
to reduce the turnpike phenomenon, we construct terminal constraints for the states $\bar{\mathbf{x}}(N)$ \eqref{eq:FHOCP:fin}.
In Fig.~\ref{fig:TCcomparisona}-\ref{fig:TCcomparisonb}, we show the optimal trajectory of the nominal states $\bar{K}$ and $\bar{v}_{\mathrm{m},1}$, i.e., the first and third components of the nominal state $\bar{x}$, over one prediction horizon, with and without terminal conditions.
From this, it can be seen that the turnpike effect is significantly reduced.
Similar results can be found for $x_{\mathrm{m}1}$,$x_{\mathrm{m}2}$ and $v_{\mathrm{m}2}$. 

Apart from the reduction in the turnpike behavior over the prediction horizon, in Fig.~\ref{fig:TCcomparisonc}-\ref{fig:TCcomparisond} we show how the inclusion of the terminal constraint \eqref{eq:FHOCP:fin} improves the controller performance.
Indeed, in Fig.~\ref{fig:TCcomparisonc} we show that, at above rated wind speed $v_{\mathrm{w}}=16$ m/s, the $P_{\mathrm{g}}$ resulting from the controller with a terminal constraint is  larger than the case without \eqref{eq:FHOCP:fin}. 
Furthermore, in Fig.~\ref{fig:TCcomparisond} we see that there is a reduction of  computation time for the solution of the FHOCP with the terminal constraint compared to the FHOCP without \eqref{eq:FHOCP:fin}, across a wide range of wind speeds.

   %\begin{figure}[thpb]
    %  \centering
    %  \framebox{
    %  %\includegraphics[scale=0.4]{figs/ModeShape.jpg}
    %  }
    %  \caption{Performance comparison among with two modes damping activated, without tower damping activated and one mode damping control activated}
    %  \label{PerformancComparison}
  % \end{figure}

\section{Conclusions} \label{sec:conclusions}
\noindent In this paper, we develop a load-limiting control based on the economic model predictive control (eMPC) framework to mitigate tower fatigue loads on multiple locations of wind turbines (WTs). 
In detail, a multi-mode vibration model is first incorporated into the eMPC framework, which allows for load reduction on multiple tower locations. 
Secondly, the rotational dynamics of the WT are written in energy terms, allowing to obtain a linear eMPC problem with convex constraints. Finally, a simulation study is performed to illustrate the scheme effectiveness, and analyse the influence of different prediction horizons and of terminal constraints.
To evaluate the control performance of the proposed approach, also eMPC-based controllers without tower damping and with one mode damping only are implemented for comparison.
Simulation results show that the proposed controller is able to effectively reduce the vibration at multiple locations, without significant effects on the power generation. 
In addition, the terminal constraints in the eMPC framework shows good effectiveness in alleviating the turnpike behavior. 

In future work, we will further verify the proposed controller using a high-fidelity WT model, such as Fatigue, Aerodynamics, Structures, and Turbulence (FAST). \textcolor{zx}{Furthermore, the current controller can be extended for floating WTs considering tower model with multiple degrees of freedom.}

\bibliographystyle{IEEEtran}
\bibliography{references}

\begin{thebibliography}{10}
\providecommand{\url}[1]{#1}
\csname url@samestyle\endcsname
\providecommand{\newblock}{\relax}
\providecommand{\bibinfo}[2]{#2}
\providecommand{\BIBentrySTDinterwordspacing}{\spaceskip=0pt\relax}
\providecommand{\BIBentryALTinterwordstretchfactor}{4}
\providecommand{\BIBentryALTinterwordspacing}{\spaceskip=\fontdimen2\font plus
\BIBentryALTinterwordstretchfactor\fontdimen3\font minus
  \fontdimen4\font\relax}
\providecommand{\BIBforeignlanguage}[2]{{%
\expandafter\ifx\csname l@#1\endcsname\relax
\typeout{** WARNING: IEEEtran.bst: No hyphenation pattern has been}%
\typeout{** loaded for the language `#1'. Using the pattern for}%
\typeout{** the default language instead.}%
\else
\language=\csname l@#1\endcsname
\fi
#2}}
\providecommand{\BIBdecl}{\relax}
\BIBdecl

\bibitem{council2021gwec}
``Global wind report 2021,'' {Global Wind Energy Council}, Tech. Rep., 2021.

\bibitem{Mulders2020}
S.~P. Mulders, T.~G. Hovgaard, J.~D. Grunnet, and J.-W. van Wingerden,
  ``Preventing wind turbine tower natural frequency excitation with a quasi-lpv
  model predictive control scheme,'' \emph{Wind Energy}, vol.~23, no.~3, pp.
  627--644, 2020.

\bibitem{Lara2021}
M.~Lara, J.~Garrido, M.~L. Ruz, and F.~Vázquez, ``Adaptive pitch controller of
  a large-scale wind turbine using multi-objective optimization,''
  \emph{Applied Sciences}, vol.~11, no.~6, 2021.

\bibitem{Shaltout2017}
M.~L. Shaltout, Z.~Ma, and D.~Chen, ``{An Adaptive Economic Model Predictive
  Control Approach for Wind Turbines},'' \emph{Journal of Dynamic Systems,
  Measurement, and Control}, vol. 140, no.~5, 12 2017, 051007.

\bibitem{AtinJoP2022}
A.~Pamososuryo, Y.~Liu, T.~Hovgaard, R.~Ferrari, and J.~{Van Wingerden},
  ``\BIBforeignlanguage{English}{Individual pitch control by convex economic
  model predictive control for wind turbine side-side tower load
  alleviation},'' \emph{\BIBforeignlanguage{English}{Journal of Physics:
  Conference Series}}, vol. 2265, no.~3, 2022, 2022 Science of Making Torque
  from Wind, TORQUE 2022.

\bibitem{rawlings2012fundamentals}
J.~B. Rawlings, D.~Angeli, and C.~N. Bates, ``Fundamentals of economic model
  predictive control,'' in \emph{2012 IEEE 51st IEEE conference on decision and
  control (CDC)}.\hskip 1em plus 0.5em minus 0.4em\relax IEEE, 2012, pp.
  3851--3861.

\bibitem{faulwasser2018economic}
T.~Faulwasser, L.~Gr{\"u}ne, M.~A. M{\"u}ller \emph{et~al.}, ``Economic
  nonlinear model predictive control,'' \emph{Foundations and
  Trends{\textregistered} in Systems and Control}, vol.~5, no.~1, pp. 1--98,
  2018.

\bibitem{gros2013economic}
S.~Gros, ``An economic nmpc formulation for wind turbine control,'' in
  \emph{52nd IEEE Conference on Decision and Control}.\hskip 1em plus 0.5em
  minus 0.4em\relax IEEE, 2013, pp. 1001--1006.

\bibitem{gawronskiAdvancedStructuralDynamics2004}
W.~K. Gawronski, \emph{Advanced Structural Dynamics and Active Control of
  Structures}, ser. Mechanical Engineering Series.\hskip 1em plus 0.5em minus
  0.4em\relax {New York}: {Springer-Verlag}, 2004.

\bibitem{jonkmanDefinition5MWReference2009}
J.~Jonkman, S.~Butterfield, W.~Musial, and G.~Scott, ``Definition of a 5-{{MW
  Reference Wind Turbine}} for {{Offshore System Development}},'' Tech. Rep.
  NREL/TP-500-38060, 947422, Feb. 2009.

\bibitem{Hovgaard2015}
T.~G. Hovgaard, S.~Boyd, and J.~B. Jørgensen, ``Model predictive control for
  wind power gradients,'' \emph{Wind Energy}, vol.~18, no.~6, pp. 991--1006,
  2015.

\bibitem{ruiterkampModellingControlLaterala}
G.~G.~J. Ruiterkamp, ``\BIBforeignlanguage{english}{Modelling and {{Control}}
  of {{Lateral Wind Turbine Tower Dynamics}}},'' Master's thesis, Delft
  University of Technology, 2021.

\bibitem{Zhang7117746}
Z.~Zhang, S.~R.~K. Nielsen, F.~Blaabjerg, and D.~Zhou, ``Dynamics and control
  of lateral tower vibrations in offshore wind turbines by means of active
  generator torque,'' \emph{Energies}, vol.~7, no.~11, pp. 7746--7772, 2014.

\bibitem{Branlard2019}
E.~S.~P. Branlard, ``Flexible multibody dynamics using joint coordinates and
  the rayleigh-ritz approximation: The general framework behind and beyond
  flex,'' \emph{Wind Energy}, vol.~22, no.~7, pp. 877--893, 2019.

\bibitem{SchlipfLIDAR2013}
D.~Schlipf, D.~J. Schlipf, and M.~Kühn, ``Nonlinear model predictive control
  of wind turbines using lidar,'' \emph{Wind Energy}, vol.~16, no.~7, pp.
  1107--1129, 2013.

\bibitem{boyd2004convex}
S.~Boyd, S.~P. Boyd, and L.~Vandenberghe, \emph{Convex optimization}.\hskip 1em
  plus 0.5em minus 0.4em\relax Cambridge university press, 2004.

\end{thebibliography}

% \begin{thebibliography}{99}
% \bibitem{c1} Global Wind Energy Council, “Global wind report 2021,” Global wind energy council, Report, 2021.
% \end{thebibliography}

\end{document}